\documentclass[twocolumn,aps,prb,final,amsfonts,amssymb,amsmath,%
floatfix,superscriptaddress,byrevtex]{revtex4}

\usepackage[]{graphicx}
\usepackage{bm}

\newcommand{\eg}{e.g., }

\begin{document}

\title{Time- and Spectrally-Resolved PL Study of a Regular Array
of InP/InAs/InP Core-multishell Nanowires}

\author{B.~Pal}
\email[E-mail: ]{bipulpal@sakura.cc.tsukuba.ac.jp}
\affiliation{Institute of Physics, University of Tsukuba, Tsukuba
305-8571, Japan}

\author{K.~Goto}
\affiliation{Institute of Physics, University of Tsukuba, Tsukuba
305-8571, Japan}

\author{M.~Ikezawa}
\affiliation{Institute of Physics, University of Tsukuba, Tsukuba
305-8571, Japan}

\author{Y.~Masumoto}
\affiliation{Institute of Physics, University of Tsukuba, Tsukuba
305-8571, Japan}

\author{P.~Mohan}
\affiliation{Research Center for Integrated Quantum Electronics,
Hokkaido University, Sapporo 060-8628, Japan}

\author{J.~Motohisa}
\affiliation{Research Center for Integrated Quantum Electronics,
Hokkaido University, Sapporo 060-8628, Japan}

\author{T.~Fukui}
\affiliation{Research Center for Integrated Quantum Electronics,
Hokkaido University, Sapporo 060-8628, Japan}

\date{\today}

\begin{abstract}
Time- and spectrally-resolved PL from a periodic array of
InP/InAs/InP core-multishell nanowires is presented. InAs layer
shows multipeak PL spectra. PL decay is nonexponential and very
slow, with decay rate depending on energy.
\end{abstract}

\maketitle

Nanometer-scale semiconductor heterostructures such as quantum
dots (0D), quantum wires (1D), and quantum wells (2D) have been
interesting research target due to their unique size dependent
electronic and optical properties associated with the lower
dimensionality and quantum confined effect. There have been
revived interests in the semiconductor nanowires due to the recent
success in the growth and fabrication of regular array of
core-shell and core-multishell nanowires
(CMNs).~\cite{lauhonnature420,mohanapl88} Such nanowires can be
used as building blocks of sophisticated nanoscale electronic and
photonic devices because of their capability to function as both
device element as well as the interconnecting wires for the
devices. Incorporation of heterostructures into the nanowires
opens up the possibility to build multifunctionality into a single
device. Though much efforts have been devoted to the growth and
fabrication of periodic array of uniformly grown CMNs, carrier
dynamics in such structures has not been studied so far by
time-resolved optical spectroscopy.

In this paper we report our study of time- and spectrally-resolved
photoluminescence (PL) from a regular array of InP/InAs/InP CMNs,
in which the InAs layer acts as a strained quantum well (QW)
embedded in the higher bandgap InP material. PL spectra from the
InAs layer show multiple peaks due to monolayer (ML) scale
variation in the layer thickness over the CMN array. Decay rate
measured within a PL peak is found to be energy dependent,
suggesting the presence of spectral diffusion due to inhomogeneous
broadening. The PL decays nonexponentially showing very slowly
decaying component surviving beyond 100~ns. This behavior may be
resulted from the spreading of the electron wavefunction of the
very thin InAs layer into the InP barriers.

\begin{figure}[tbh]
\includegraphics[clip,width=7.7cm]{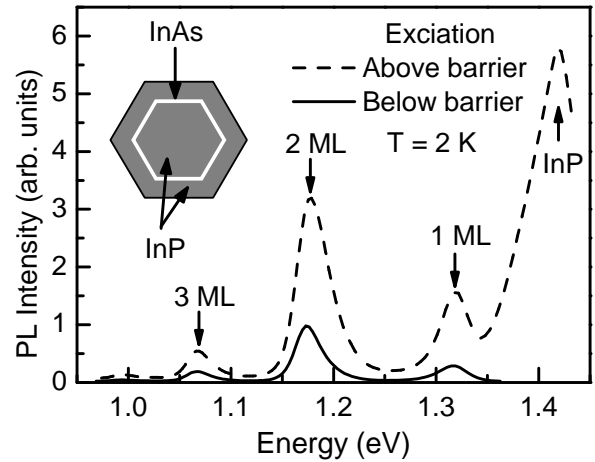}
\caption{\label{fig1} Schematic cross-section of a nanowire and PL
spectra of the InAs layer for excitation above and below the InP
barrier.}
\end{figure}

The sample studied here is a periodic array of uniformly grown,
vertically oriented InP/InAs/InP CMNs grown by using selective
area metalorganic vapor phase epitaxy. Each nanowire has a
hexagonal-cylindrical symmetry and it consists of an inner InP
core and InAs and InP inner and outer shells, respectively, in
which the InP core and the InP outer shell serve as the barrier
layers while the InAs shell acts as a strained QW layer [see
Fig.~\ref{fig1}-inset]. Further details of the sample structure
and growth procedure may be found in Ref.~\onlinecite{mohanapl88}.

The PL spectra from this sample are measured at 2~K for excitation
above and below the InP barrier by a Ti:Sapphire laser. A
monochromator with a spectral resolution of about 1~meV, a liquid
$N_{2}$-cooled InGaAsP photomultiplier tube, and a photon counter
is used. The focused laser spot size on the sample was about
150~$\mu$m in diameter covering about 3000 nanowires. The highest
energy peak observed in Fig.~\ref{fig1} for above barrier
excitation is associated with the InP barrier. The remaining peaks
observed for both above and below barrier excitation can be
assigned to the InAs QWs. The spectra are measured from an
ensemble of CMNs. In the ensemble, possibly there are monolayer
scale variations in the thickness of InAs QW layer. From a
calculation of the ground-state transition energy in strained InAs
QWs with InP barriers, the main three peaks are assigned to 1~ML,
2~ML, and 3~ML wide QWs, respectively, from higher to lower
energy.~\cite{mohanapl88}

%%%%%%%%%%%%%%%%%%%%%%%%%%%%%%%%%%%%%%%%%%%%%%%%%%%%%%%%%%%%
%%%%%%%%%%%%%%%%%%%%%%%%%%%%%%%%%%%%%%%%%%%%%%%%%%%%%%%%%%%%

\begin{figure}[tbh]
\includegraphics[clip,width=7.85cm]{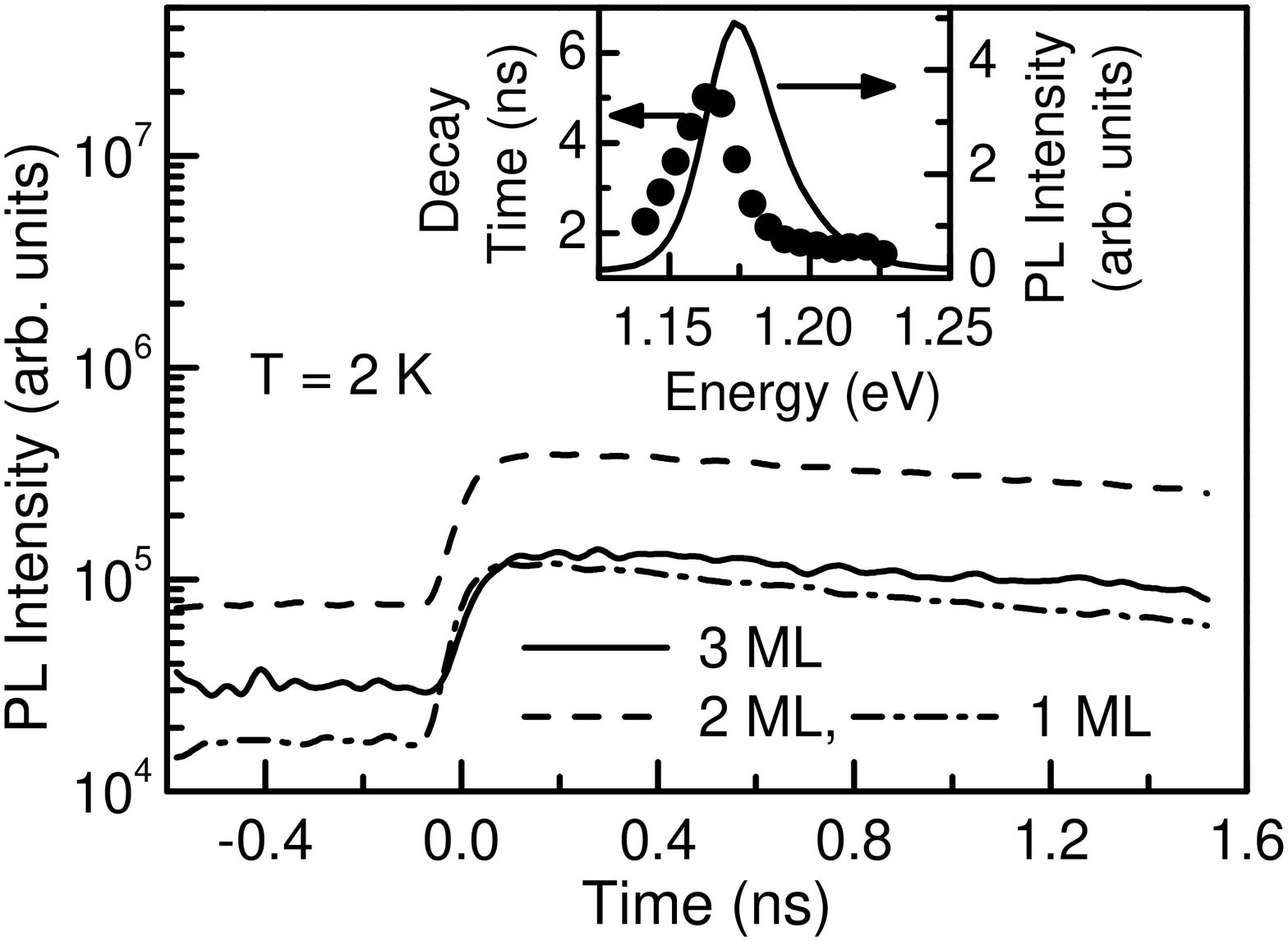}
\caption{\label{fig2} PL decay at all three peaks of InAs QWs.
Inset: Energy dependence of the PL decay time for the 2 ML peak.}
\end{figure}

The PL decay for the InAs peaks is measured over a 2~ns time-span
by using a synchroscan infrared streak camera and a picosecond
Ti:Sapphire laser with 82~MHz repetition rate. The data is plotted
in Fig.~\ref{fig2}, showing very similar decay rates for all three
peaks. In the 2~ns time-span the decay looks like an exponential,
giving about 4~ns decay time. However, presence of a large PL
signal at negative times before the arrival of the excitation
pulse suggests that very long decay components are present.

%%%%%%%%%%%%%%%%%%%%%%%%%%%%%%%%%%%%%%%%%%%%%%%%%%%%%%%%%%%%
%%%%%%%%%%%%%%%%%%%%%%%%%%%%%%%%%%%%%%%%%%%%%%%%%%%%%%%%%%%%

\begin{figure}[tbh]
\includegraphics[clip,width=8.3cm]{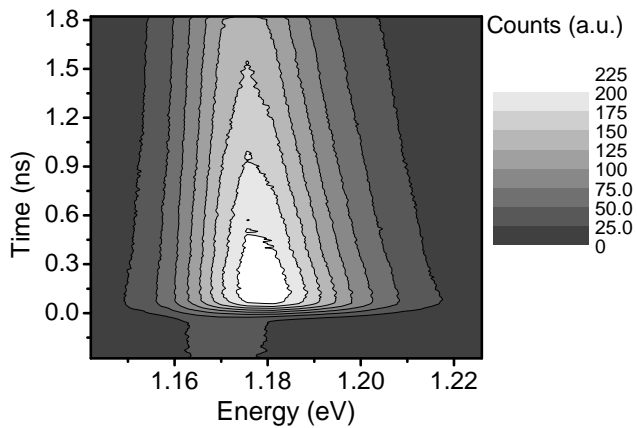}
\caption{\label{fig3} Two-dimensional plot of PL intensity
(grayscale coded) on the time-energy surfaces.}
\end{figure}

We find that for a given peak, the decay rate depends on the
spectral energy. We systematically measure the PL decay at
different spectral energy within the 2~ML peak. A dependence of PL
decay time on the spectral position is shown in the inset of
Fig.~\ref{fig2}. A two-dimensional plot of the grayscale coded PL
intensity on to the time-energy surface for the 2~ML peak in
Fig.~\ref{fig3} shows that the spectral weight is shifted towards
the lower energy at increased time. This suggests the presence of
spectral diffusion due the inhomogeneous broadening present in the
sample. This may be resulted from the inhomogeneous strain in the
QWs in the ensemble of CMNs.~\cite{mohanapl88}

%%%%%%%%%%%%%%%%%%%%%%%%%%%%%%%%%%%%%%%%%%%%%%%%%%%%%%%%%%%%
%%%%%%%%%%%%%%%%%%%%%%%%%%%%%%%%%%%%%%%%%%%%%%%%%%%%%%%%%%%%

\begin{figure}[tbh]
\includegraphics[clip,width=7.9cm]{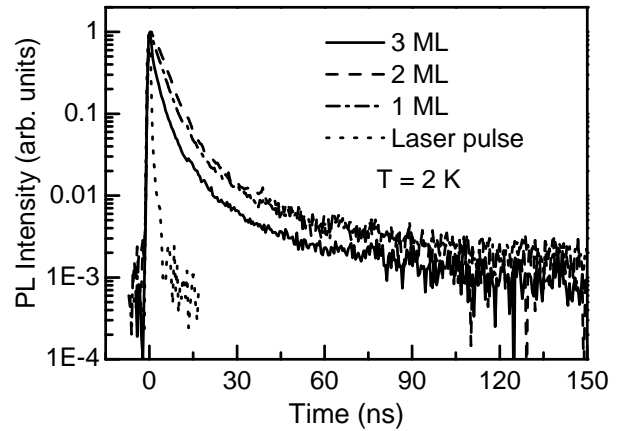}
\caption{\label{fig4} Nonexponential PL decay extending over 100
ns. Measured laser pulse profile is also shown.}
\end{figure}

To investigate the PL decay over a longer time-span we reduce the
laser repetition rate to 800kHz by using a pulse picker. Then the
PL decay is measured by using the time-correlated single photon
counting technique. Figure~\ref{fig4} shows the PL decay over a
200~ns time-span for the 1, 2, and 3~ML peaks of the InAs QWs. The
measured time-profile of the laser pulse is also shown, giving a
time resolution of about 1~ns. The PL decay is nonexponential and
very long components surviving beyond 100~ns are present.

%%%%%%%%%%%%%%%%%%%%%%%%%%%%%%%%%%%%%%%%%%%%%%%%%%%%%%%%%%%%
%%%%%%%%%%%%%%%%%%%%%%%%%%%%%%%%%%%%%%%%%%%%%%%%%%%%%%%%%%%%

To understand the origin of the slow decay of PL we may consider
the following. For very thin InAs QWs, the electron state is not
very well confined and there may be appreciable spreading and
penetration of the QW electron wavefunction into the barrier
layer.~\cite{cebullaprb39} In a way, the PL can be considered as
the indirect transition in a type-II material. In such a case,
slow and nonexponential decay of PL is reported in the
literature.~\cite{minamiprb36} The slow decay is a result of
insufficient overlap of the electron and hole wavefunctions. The
nonexponential decay results from a distribution of decay rates.
Further study by theoretical modelling and experiments to measure
the temperature dependence of the decay rate and PL intensity is
in progress to clarify the issue.

In summery, time- and spectrally-resolved PL from a periodic array
of InP/InAs/InP core-multishell nanowires is presented. InAs layer
shows multipeak PL spectra. PL decay is nonexponential and very
slow, with decay rate depending on energy.

\end{document}